\documentclass[conference]{IEEEtran}
\IEEEoverridecommandlockouts
\usepackage{cite}
\usepackage{amsmath,amssymb,amsfonts}
\usepackage{float}
\usepackage[para,online,flushleft]{threeparttable}
\usepackage{booktabs}
\usepackage{multirow}
\usepackage{newtxmath}
\usepackage{tabularx}
\usepackage{pifont}
\makeatletter
\usepackage{rotating}
\makeatother
\usepackage{color, colortbl}
\usepackage{graphicx}
\usepackage{ragged2e}
\usepackage{textcomp}
\usepackage{lettrine}
\usepackage{xcolor}
\usepackage{hyperref}
\usepackage{algpseudocode}
\usepackage{algorithm}
\usepackage{acronym}
\usepackage{tabu}
\usepackage{authblk}
\usepackage[english]{babel}
\usepackage{blindtext}
\usepackage{makecell}
\usepackage{rotating}
\usepackage{multirow}
\usepackage{cellspace}
\usepackage{orcidlink}
\acrodef{AI}[AI]{Artificial Intelligence}
\acrodef{AIMC}[AIMC]{Analog-Based In-Memory Computing}
\acrodef{NLP}[NLP]{Natural Language Processing}
\acrodef{DNN}[DNN]{Deep Neural Network}
\acrodef{VMM}[VMM]{Vector-Matrix Multiplication}
\acrodef{HWA}[HWA]{Hardware-Aware}
\acrodef{DAC}[DAC]{Digital-to-Analog Converter}
\acrodef{LRA}[LRA]{Long-Range Arena}
\acrodef{ADC}[ADC]{Analog-to-Digital Converter}
\acrodef{SNR}[SNR]{Signal-to-Noise Resolution}
\acrodef{FP}[FP]{Floating-Point}
\acrodef{QAT}[QAT]{Quantization-Aware Training}
\acrodef{PTQ}[PTQ]{Post-Training Quantization}
\acrodef{GPU}[GPU]{Graphics Processor Unit}
\acrodef{PE}[PE]{Processing Element}
\acrodef{WL}[WL]{Word-Line}
\acrodef{BL}[BL]{Bit-Line}
\acrodef{SOTA}[SOTA]{State-Of-the-Art}
\acrodef{AIHWKIT}[AIHWKIT]{IBM Analog Hardware Acceleration Kit}
\acrodef{PCM}[PCM]{Phase-Change Memory}
\acrodef{LDPU}[LDPU]{Local Digital Processing Unit}
\acrodef{PWM}[PWM]{Pulse-Width Modulation}
\acrodef{MVM}[MVM]{Matrix-Vector Matrix}
\acrodef{GLUE}[GLUE]{General Language Understanding Evaluation}
\acrodef{PT}{Post-Training}
\newcolumntype{R}{>{\raggedleft\arraybackslash}X}
\newcommand{\analog}[1]{\breve{#1}}
\algnewcommand\algorithmicforeach{\textbf{for each}}
\algdef{S}[FOR]{ForEach}[1]{\algorithmicforeach\ #1\ \algorithmicdo}
\begin{document}
\title{Improving the Accuracy of Analog-Based In-Memory Computing Accelerators Post-Training}
\author[1,$\dagger$]{Corey Lammie\orcidlink{0000-0001-5564-1356}\thanks{\hspace{-1em}\rule{3cm}{0.5pt} \newline \textcopyright  \hspace{1pt} 2024 IEEE. Personal use of this material is permitted. Permission from IEEE must be obtained for all other uses, in any current or future media, including reprinting/republishing this material for advertising or promotional purposes, creating new collective works, for resale or redistribution to servers or lists, or reuse of any copyrighted component of this work in other works.}}
\author[1]{Athanasios Vasilopoulos\orcidlink{0009-0001-9081-6139}}
\author[1,2]{Julian Büchel\orcidlink{0000-0001-9495-7150}}
\author[1,2]{Giacomo Camposampiero}
\author[1]{\authorcr Manuel Le Gallo\orcidlink{0000-0003-1600-6151}}
\author[1]{Malte Rasch\orcidlink{0000-0002-7988-4624}}
\author[1,$\dagger$]{Abu Sebastian\orcidlink{0000-0001-5603-5243}}

\affil[1]{IBM Research Europe, 8803 Rüschlikon, Switzerland} % \authorcr %Email:\{corey.lammie, mostafa.rahimiazghadi\}@jcu.edu.au}
\affil[2]{Department of Computer Science, ETH Zürich, Switzerland}
\affil[$\dagger$]{Corresponding authors. Emails: \{corey.lammie@ibm.com, ase@zurich.ibm.com\}}

\maketitle
\begin{abstract}
\ac{AIMC} inference accelerators can be used to efficiently execute \ac{DNN} inference workloads. However, to mitigate accuracy losses, due to circuit and device non-idealities, \ac{HWA} training methodologies must be employed. These typically require significant information about the underlying hardware. 
In this paper, we propose two \ac{PT} optimization methods to improve accuracy after training is performed.
For each crossbar, the first optimizes the conductance range of each column, and the second optimizes the input, i.e, \ac{DAC}, range.
It is demonstrated that, when these methods are employed, the complexity during training, and the amount of information about the underlying hardware can be reduced, with no notable change in accuracy ($\leq$0.1\%) when finetuning the pretrained RoBERTa transformer model for all \ac{GLUE} benchmark tasks. Additionally, it is demonstrated that further optimizing learned parameters \ac{PT} improves accuracy.
\end{abstract}


%------------------------------------------------------------
% INTRODUCTION
%------------------------------------------------------------
\section{Introduction}
\acresetall
\lettrine{A}{rtificial} Intelligence (AI) has been successfully applied across diverse domains to solve challenging engineering problems, ranging from image classification to \ac{NLP}~\cite{chellappaAdvancesMachineLearning2021}. As the complexity of AI models, in particular, \acp{DNN}, continues to increase, dedicated accelerators are required to accelerate the inference workloads of these models in resource-contained environments~\cite{xuScalingEdgeInference2018,rayReviewTinyMLStateoftheart2022,azghadiHardwareImplementationDeep2020,legallo64coreMixedsignalInmemory2023a}. \ac{AIMC} accelerators are one such type of accelerator, which have gained significant interest, due to their ability to execute \acp{VMM} in $\mathcal{O}(1)$ time-complexity~\cite{lammieVariationAwareBinarizedMemristive2019,sebastianMemoryDevicesApplications2020a,songRecentAdvancesFuture2023}. 
While there are significant efforts to mitigate both circuit and device non-idealities~\cite{molasInvitedResistiveMemories2018,yangResearchProgressMemristor2022,buchelGradientDescentbasedProgramming2022,JETCAS,vasilopoulosExploitingStateDependency2023,liOptimizationProjectedPhase2023}, networks trained to be deployed on traditional compute hardware with \ac{FP}-precision parameters, e.g., \acp{GPU}, when deployed on \ac{AIMC} hardware, require retraining with some degree of consideration about the underlying hardware to achieve near- or at-ISO accuracy.

While some \ac{HWA} training techniques, such as \ac{QAT}, are widely adopted~\cite{parkValueawareQuantizationTraining2018}, due to the proliferation of reduced precision digital accelerators and deterministic execution flows, others require instance specific information~\cite{raschHardwareawareTrainingLargescale2023a}. Hence, they cannot easily be generalized for different hardware architectures.
Moreover, when \acp{PE} are arranged in crossbar arrangements, as done in \ac{AIMC} architectures, they require input and output scales, in addition to clipping parameters, to be specified. This usually requires the \emph{tiling} of layer weights, which further increases training complexity~\cite{raschHardwareawareTrainingLargescale2023a}.
For large models, which require significant resources and time to retrain, this is problematic.

\begin{figure}[!t]
\centering
\includegraphics[width=0.5\textwidth]{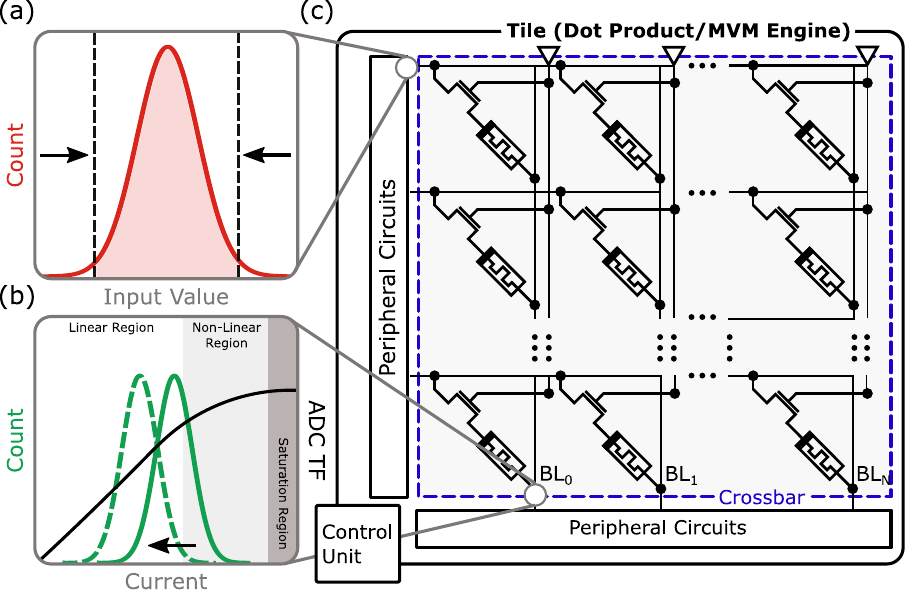}
\caption{
\textbf{By performing \ac{PT} optimization methods, accuracy can be improved}. \textbf{(a)} By reducing the input (\ac{DAC}) range, the effective dynamic resolution of \acp{ADC} can be increased, at the cost of clipping inputs. \textbf{(b)} The maximum conductance range of each column can be reduced to avoid \ac{ADC} saturation, and increased to boost the signal, improving the \ac{SNR}.
}\label{fig:concept}
\vspace{-1em}
\end{figure}

Recently, for these models, \ac{PTQ}~\cite{nagelAdaptiveRoundingPostTraining2020} methods, which quantize network parameters \ac{PT}, without the requirements of retraining or fine-tuning using training data, have burgeoned in popularity~\cite{xiaoSmoothQuantAccurateEfficient2023}.
Taking inspiration from this approach, in this paper we introduce two methods to improve the inference accuracy of \ac{AIMC} accelerators (Fig.~\ref{fig:concept}). Both employ optimization methods \ac{PT}, to determine parameters which are normally learned. The first optimizes the input (\ac{DAC}) range for all \acp{WL}, and the second optimizes the conductance range of each \ac{BL}.
Our specific contributions are as follows:
\begin{enumerate}
    \item We propose two novel \ac{PT} optimization methods to improve the accuracy of \ac{AIMC} inference accelerators \ac{PT}, which reduce training complexity;
    \item We model hardware implementations of the RoBERTa transformer model, and demonstrate, that, after finetuning, both methods can improve scores across \ac{GLUE} tasks when input ranges and conductance ranges are not learned;
    \item We establish baseline GLUE scores for the RoBERTa model on simulated \ac{AIMC}-hardware.
\end{enumerate}

\section{Preliminaries}~\label{sec:preliminaries}
% In this section, key-concepts of \ac{AIMC} will be introduced. Additionally, in the context of \ac{AIMC}, \ac{HWA} training approaches will be over-viewed. Finally, the \ac{LRA} benchmark for transformers will be described.
\vspace{-2em}
\subsection{Analog-Based In-Memory Computing}
\ac{AIMC} accelerators are capable of performing \ac{VMM} operations $\mathbf{Y}^T = \mathbf{X}^T\mathbf{W}$ using the laws of physics,  where $\mathbf{W}$ is an $M \times N$ matrix, $\mathbf{X}$ is a  $M \times 1$ vector, and $\mathbf{Y}$ is a $N \times 1$ vector.
This is done by encoding elements of $\mathbf{X}$ as \ac{WL} voltages, denoted using $\mathbf{V}$, and elements of $\mathbf{W}$ as conductances of the unit cells, denoted using $\mathbf{G}$.
The analog computation, i.e., $\mathbf{I}=\mathbf{VG}$ is performed, where the current flow to the end of the $N$-th column is $I_N = \sum_{i=0}^MG_{i,N}V_i$. By presenting rows of a matrix as $\mathbf{X}$, sequentially, or in parallel (using multiple crossbars), matrix multiplication, the most dominant operation in \acp{DNN}, can be performed. In traditional \ac{AIMC} architectures, other, i.e., \emph{auxiliary} operations, can be performed using digital compute blocks - usually at reduced fixed-point precision.

\subsection{Hardware-Aware Training}
For current \ac{AIMC} hardware, \ac{HWA} training is usually required to mitigate degradation in accuracy. To perform \ac{HWA} training, the IBM Analog Hardware Acceleration Kit (AIHWKIT)~\cite{raschFlexibleFastPyTorch2021a,galloUsingIBMAnalog2023} is used. Additional operations, including weight noise injection, are added during forward propagation passes. Mathematically, the generally simulated \ac{AIMC} forward passes can be expressed as
\begin{equation}
  \label{eq:mat-vec-analog}
    y_i = \alpha^\text{out}_i f_\text{adc}\big(\sum_j(\analog{w}_{ij} +
      \sigma_\text{w}\xi_{ij})(f_\text{dac}(x_j) 
      + \sigma_\text{inp}\xi_j)  + \sigma_\text{out}\xi_i\big) + \beta_i,
\end{equation}
where $f_\text{adc}$ and $f_\text{dac}$ model the non-linear analog-to-digital and digital-to-analog processes, with dynamic scaling and range clipping, and $\xi$ denotes Gaussian noise~\cite{raschHardwareawareTrainingLargescale2023a}. In addition to training noise-resilient weights, it is also possible to train other parameters, such as the input (\ac{DAC}) bounds of each crossbar, using back-propagation (see Section~\ref{sec:methods}). %This will be discussed in greater detail in Section~\ref{sec:methods}.

\subsection{The GLUE Benchmark and RoBERTA}
The \ac{GLUE} benchmark is a collection of resources for training, evaluating, and analyzing natural language understanding systems~\cite{wangGLUEMultiTaskBenchmark2019}. In this paper, we consider eight sentence- and sentence-pair language understanding tasks. The WNLI task is excluded, as in the original BERT study, as it contains a large number of  adversarial examples~\cite{devlinBERTPretrainingDeep2019}. We adopt the base RoBERTA model, which is a ubiquitous bidirectional transformer based on BERT~\cite{devlinBERTPretrainingDeep2019}, with approximately 125M learnable parameters~\cite{liuRoBERTaRobustlyOptimized2019}.

\begin{figure}[!t]
\centering
\includegraphics[width=0.45\textwidth]{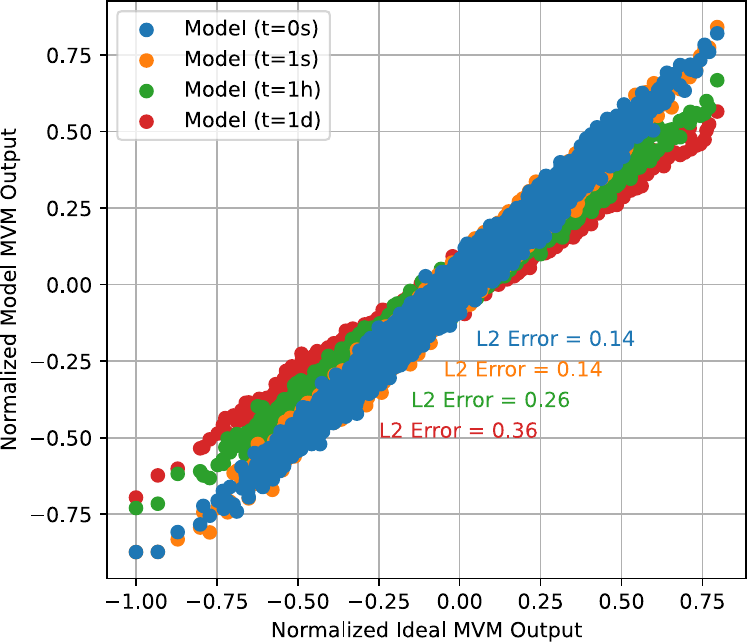}
\caption{
\textbf{The simulated and ideal \ac{MVM} outputs for 5,120 inputs} sampled from a uniform distribution, with clipped normally distributed weights ($\sigma$=0.25). At $t=0$, the L2 error is $\approx$14\%.
}\label{fig:experiment_match}
\end{figure}

\begin{table}[!b]
\centering

\begin{threeparttable}
\caption{Network utilization information and baseline task scores.}\label{table:architecture}
\begin{tabularx}{0.5\textwidth}{llR} \toprule \toprule
\multicolumn{3}{c}{\textbf{Network Utilization}} \\ \midrule
\multicolumn{2}{l}{Total number
  of network parameters} & 124,647,170 \\
\multicolumn{2}{l}{Number
  of mapped parameters} & 85,609,730 \\ 
\multicolumn{2}{l}{Number
  of tiles} & 486 \\ 
\multicolumn{2}{l}{Avg. tile utilization} & \multicolumn{1}{r}{61.57\%} \\ \addlinespace[0.5ex] \toprule \toprule
\textbf{Task} & \textbf{Metric(s)} & \textbf{Baseline (FP) Score\tnote{1} (/100)} \\ \midrule
MNLI & Accuracy & 86.2\\
QNLI & Accuracy & 92.1\\
QQP & Accuracy/F1 & 89.2/86.1\\
RTE & Accuracy & 79.4\\
SST-2 & Accuracy & 94.0\\
MRPC & Accuracy/F1 & 90.2/92.9\\
CoLA & Matthews Correlation & 63.3\\
STS-B & Pearson/Spearman Correlation & 90.3/90.0\\ \bottomrule \bottomrule
\end{tabularx}
% \begin{tabularx}{0.5\textwidth}{l|r|ccccc} \toprule \toprule
% \multicolumn{2}{c|}{\multirow{2}{*}{\textbf{Architecture}}\tnote{1}} & \multicolumn{5}{c}{\textbf{Task}} \\
% \multicolumn{2}{l|}{} & \textbf{ListOps} & \textbf{Text} & \textbf{Retrieval} & \textbf{Image} & \textbf{PF\tnote{2}} \\ \midrule
% \multirow{4}{*}{Encoder} & \begin{tabular}[c]{@{}r@{}}Attention~\\Heads\end{tabular} & 2 & 2 & 2 & 2 & 8 \\
%  & \begin{tabular}[c]{@{}r@{}}Embed~\\Dim.\end{tabular} & 64 & 64 & 64 & 64 & 128 \\
%  & \begin{tabular}[c]{@{}r@{}}FFN Embed~\\Dim.\end{tabular} & 128 & 128 & 64 & 128 & 128 \\
%  & Layers & 2 & 2 & 2 & 2 &  \\
% \multirow{2}{*}{Classifier} & Out Dim. & 128 & 128 & 512 & 128 & 200 \\
%  & \begin{tabular}[c]{@{}r@{}}Activation~\\Function\end{tabular} & gelu & silu & gelu & silu &  silu \\ \midrule
% \multicolumn{2}{c|}{\textbf{Accuracy (\%)}\tnote{1}} & 39.95 & 66.56 & 78.72 & 46.62 & 69.94 \\ \bottomrule \bottomrule
% \end{tabularx}
\begin{tablenotes}
\item[1] The pre-trained RoBERTA network was fine-tuned (without noise) and evaluated in \ac{FP}.
\end{tablenotes}
\end{threeparttable}
\end{table}

\section{Hardware Model}~\label{sec:model}
To perform realistic hardware simulations, we used an experimentally verified model, calibrated based on extensive measurements performed on an array containing 1 million \ac{PCM} devices~\cite{joshiAccurateDeepNeural2020}. A tile size of 512$\times$512 is assumed. Inputs are encoded using 8-bit \ac{PWM}, and weights are represented using a standard differential weight mapping scheme with $G_{max}=25\mu S$. Weight (programming), read, drift, and output noise is captured, in addition to IR drop with 0.35$\Omega$ wire resistance between adjacent cross-points. \acp{ADC} are assumed to operate at 8-bit precision. Saturation occurs, when for a given tile column, 10 \acp{PE} are at $G_{max}$ and the corresponding inputs are at a maximum. All auxiliary operations are assumed to be computed in \ac{FP} precision. As seen in Fig.~\ref{fig:experiment_match}, the \ac{MVM} error increases linearly with respect to logarithmic time.

\section{Methodology}~\label{sec:methods}
\vspace{-2em}

\subsection{Initial Training}
RoBERTA was originally pre-trained in \ac{FP} precision using five English-language corpora of varying sizes and domains, totaling over 160GB of uncompressed text~\cite{liuRoBERTaRobustlyOptimized2019}.

\subsection{Mapping}
Both convolutional and linear (dense) layers were unfolded and mapped to tiles using a symmetrical weight mapping scheme, where layers that spanned more than one tile were distributed evenly. The residual weights, i.e., $\text{mod}(m, n)$, where $m$ is the length of the one weight dimension and $n$ is the size of the tile along this dimension, were mapped to the last tile(s). In Table~\ref{table:architecture}, the number of mapped parameters and tiles, in addition to the tile utilization, is reported.
\subsection{Hardware-Aware Finetuning}\label{sec:hwa_retraining}
For each configuration, networks were finetuned using the mismatched validation set for the MNLI task and the validation set for all other tasks.
\ac{HWA} training was performed for 20 epochs with an initial learning rate of 5E-5 and a linearly decaying (to 5E-8) learning rate schedule. A batch size of 8 was used and the maximum sequence length was set to 128. The AdamW~\cite{loshchilovDecoupledWeightDecay2019} optimizer was used.
Unit valued weights were used to represent conductance values, which were linearly mapped to conductance values after training.
Additive noise sampled from a normal distribution with a standard deviation value of 0.06 was injected to unit weight values during forward propagation passes, and output noise was sampled from a normal distribution with a standard deviation of 0.1.
For some configurations, during finetuning, input range and channel-wise conductance range learning was performed. The additional training steps for each are described as follows.
\subsubsection{Input Range Learning}
can be used to find the optimal input range for each crossbar during \ac{HWA} training by computing additional gradients as described in Eq.~\ref{eq:input_range_learning}

\begin{algorithm}[!b]
\caption{\ac{PT} optimization methods.}\label{alg:methods}
\begin{algorithmic}[1]
\Require N. input samples, $N$, percentile $K$, N. standard deviations, $L$, maximum $G_{max}$, and minimum $G_{min}$ largest conductance values, \ac{ADC} saturation current $I_S$

\Procedure{input\_range\_optimization}{$N, K$}
  \ForEach {tile}
    \State Collect $N$ input samples, $\mathbf{X}$
    \State $\mathbf{X_r} = \text{percentile}(\mathbf{X}, K)$ \Comment{Set the input range, $\mathbf{X_r}$, to the $K$th percentile of all input samples}
  \EndFor
\EndProcedure

\Procedure{conductance\_range\_optimization}{$N$, $K$, $G_{max}$, $G_{min}$, $I_S$}
  \ForEach {tile}
     \State Collect $N$ input samples, $\mathbf{X}$
        \ForEach {\ac{BL}}
            \State $G_{\text{BL\_max}} \leftarrow G_{max}$ \Comment{Additionally, rescale all \ac{PE} conductance values, by $G_{\text{BL\_max}} / G_{\text{max}}$}     
            \State $\mathbf{I} = \mathbf{V_{\text{read}}} \mathbf{G_{\text{BL}}}[\mathbf{X}[\mathbf{X} != 0] y_{\text{factor}}$ \Comment{$y_{\text{factor}}$ is a linear correlation factor used to relate the \ac{ADC} count and conductance}
            \State $\mu = \text{mean}(\mathbf{I}), \sigma = \text{std}(\mathbf{I}), I_P = \mu + L\mathcal{N}(0,\sigma)$
            \If{$I_P >I_S$}
                \State $G_{max}$ = $\text{min(}G_{max}I_S/I_P,G_{min}$)
            \EndIf
        \EndFor
  \EndFor
\EndProcedure
\end{algorithmic}
\end{algorithm}

\begin{figure}[!t]
\centering
\includegraphics[width=0.425\textwidth]{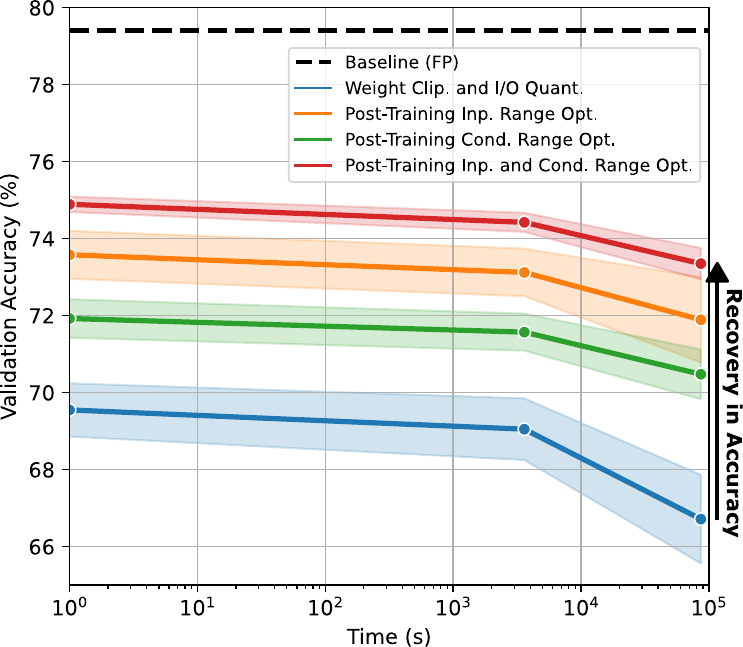}
\caption{
\textbf{Temporal dependence} of the mean and standard deviation (10 repetitions) of the validation accuracy after \ac{HWA} finetuning on the RTE task. For all three \ac{PT} configurations, weight clipping and I/O quantization was also performed. Input and conductance range learning was not performed.
}\label{fig:time_scale}
\end{figure}

\begin{equation}\label{eq:input_range_learning}
\begin{split}
    \nabla_{X_r} = \big(\sum \text{clamp}(\nabla_{X} [X \geq X_r], \text{max}=0) \\- \sum \text{clamp}(\nabla_{X} [X \leq -X_r], \text{min}=0)\big) \\+ X_r \eta (\frac{\sum \text{abs}(X < X_r)}{X.\text{numel()}}).
\end{split}
\end{equation}
$X_r$ denotes the input range, $X$ denotes cached \ac{WL} inputs, and $\eta$ is a user-specifiable constant decay term. For all relevant configurations, this was set to the default value of 1E-2.

\setlength{\tabcolsep}{2.5pt}
\begin{table*}[!t]
\centering
\begin{threeparttable}[!t]
\centering
\caption{The mean and standard deviation of all (/100) GLUE task scores at t=1h, for different configurations (10 repetitions). For each task, the best configuration(s) are bolded and underlined. For each configuration, the average score is also reported.}\label{table:main_results}
\begin{tabular}{ll|llll|cccccccc|c} \toprule \toprule
\multicolumn{6}{c|}{\textbf{Configuration}} & \multicolumn{9}{c}{\textbf{GLUE Task}} \\ \midrule
\multirow[t]{2}{*}{\rotcell[c]{\textbf{Weight Clipping}}} &
\multirow[t]{2}{*}{\rotcell[c]{\textbf{Quant I/O.}}} &
\begin{sideways}\textbf{Conductance}\end{sideways} & \begin{sideways}\textbf{Range}\end{sideways} & \begin{sideways}\textbf{Input}\end{sideways} & \begin{sideways}\textbf{Range}\end{sideways} &

\multirow[t]{2}{*}{\parbox{1.1cm}{\centering \textbf{MNLI}\\ \centering Accuracy}} & 
\multirow[t]{2}{*}{\parbox{1.1cm}{\centering \textbf{QNLI}\\ \centering Accuracy}} &
\multirow[t]{2}{*}{\parbox{2cm}{\centering \textbf{QQP}\\ \centering Accuracy/F1}} &
\multirow[t]{2}{*}{\parbox{1.1cm}{\centering \textbf{RTE}\\ \centering Accuracy}} &
\multirow[t]{2}{*}{\parbox{1.1cm}{\centering \textbf{SST-2}\\ \centering Accuracy}} &
\multirow[t]{2}{*}{\parbox{2cm}{\centering \textbf{MRPC}\\ \centering Accuracy/F1}} &
\multirow[t]{2}{*}{\parbox{1.25cm}{\centering \textbf{CoLA}\\ \centering Matthews\\ \centering Correlation}}  &
\multirow[t]{2}{*}{\parbox{2cm}{\centering \textbf{STS-B}\\ \centering Pearson/ \\ Spearman Correlation}} &
\multirow[t]{2}{*}{\textbf{Avg.}} \\ 
\cline{3-6} \addlinespace[0.5ex]
 &  & \multicolumn{1}{r}{\begin{sideways}\textbf{Learned}\end{sideways}} & \multicolumn{1}{r}{\begin{sideways}\textbf{Post-}\tnote{1}\end{sideways}} & \multicolumn{1}{|r}{\begin{sideways}\textbf{Learned}\end{sideways}} & \multicolumn{1}{r|}{\begin{sideways}\textbf{Post-}\tnote{1}\end{sideways}} &  &  &  &  &  &  &  &  &  \\ \midrule

\multicolumn{6}{c|}{\textbf{Ideal}\tnote{2}} & \cellcolor{gray!0}{85.6} & \cellcolor{gray!0}{91.7} & \cellcolor{gray!0}{88.7/85.5} & \cellcolor{gray!0}{78.9} & \cellcolor{gray!0}{93.5} & \cellcolor{gray!0}{89.9/92.3} & \cellcolor{gray!0}{62.7} & \cellcolor{gray!0}{90.0/89.9} & \cellcolor{gray!0}{\textbf{85.9}} \\
\midrule
\ding{51} & \ding{55} & \ding{55} & \ding{55} & \ding{55} & \ding{55} & \cellcolor{gray!0}{74.1(0.11)} & \cellcolor{gray!0}{79.2(0.12)} & \cellcolor{gray!0}{76.1(0.13)/74.3(0.13)} & \cellcolor{gray!0}{68.5(0.08)} & \cellcolor{gray!0}{82.3(0.13)} & \cellcolor{gray!0}{77.4(0.11)/81.4(0.14)} & \cellcolor{gray!0}{55.3(0.04)} & \cellcolor{gray!0}{78.5(0.13)/78.5(0.13)} & \cellcolor{gray!0}{\textbf{75.1}} \\
\ding{51} & \ding{51} & \ding{55} & \ding{55} & \ding{55} & \ding{55} & \cellcolor{gray!3}{74.5(0.12)} & \cellcolor{gray!2}{79.6(0.13)} & \cellcolor{gray!15}{78.4(0.12)/75.2(0.12)} & \cellcolor{gray!4}{69.0(0.10)} & \cellcolor{gray!5}{83.0(0.15)} & \cellcolor{gray!10}{78.9(0.12)/82.7(0.16)} & \cellcolor{gray!2}{55.5(0.04)} & \cellcolor{gray!8}{79.8(0.14)/82.1(0.11)} & \cellcolor{gray!8}{\textbf{76.3}} \\ \midrule
\ding{51} & \ding{51} & \ding{55} & \ding{55} & \ding{55} & \ding{51} & \cellcolor{gray!16}{76.5(0.10)} & \cellcolor{gray!28}{83.7(0.13)} & \cellcolor{gray!23}{79.6(0.12)/77.1(0.11)} & \cellcolor{gray!23}{71.6(0.10)} & \cellcolor{gray!9}{83.7(0.13)} & \cellcolor{gray!10}{79.0(0.13)/83.8(0.17)} & \cellcolor{gray!6}{56.0(0.04)} & \cellcolor{gray!18}{81.4(0.15)/82.3(0.14)} & \cellcolor{gray!18}{\textbf{77.7}} \\
\ding{51} & \ding{51} & \ding{55} & \ding{55} & \ding{51} & \ding{55} & \cellcolor{gray!35}{79.2(0.11)} & \cellcolor{gray!39}{85.5(0.16)} & \cellcolor{gray!40}{82.1(0.13)/79.8(0.12)} & \cellcolor{gray!34}{73.1(0.12)} & \cellcolor{gray!42}{89.2(0.16)} & \cellcolor{gray!40}{83.6(0.13)/87.4(0.17)} & \cellcolor{gray!42}{60.0(0.05)} & \cellcolor{gray!43}{85.3(0.17)/84.9(0.15)} & \cellcolor{gray!39}{\textbf{80.9}} \\
\ding{51} & \ding{51} & \ding{55} & \ding{55} & \ding{51} & \ding{51} & \cellcolor{gray!37}{79.5(0.09)} & \cellcolor{gray!40}{85.7(0.17)} & \cellcolor{gray!35}{81.4(0.17)/80.4(0.15)} & \cellcolor{gray!26}{72.0(0.10)} & \cellcolor{gray!36}{88.1(0.16)} & \cellcolor{gray!38}{83.3(0.12)/87.6(0.15)} & \cellcolor{gray!35}{59.3(0.05)} & \cellcolor{gray!35}{84.0(0.15)/85.2(0.13)} & \cellcolor{gray!37}{\textbf{80.6}} \\
\ding{51} & \ding{51} & \ding{55} & \ding{51} & \ding{55} & \ding{55} & \cellcolor{gray!35}{79.3(0.13)} & \cellcolor{gray!35}{84.8(0.13)} & \cellcolor{gray!34}{81.3(0.14)/79.4(0.13)} & \cellcolor{gray!34}{73.1(0.10)} & \cellcolor{gray!34}{87.9(0.14)} & \cellcolor{gray!34}{82.6(0.16)/86.9(0.17)} & \cellcolor{gray!35}{59.2(0.05)} & \cellcolor{gray!34}{83.9(0.17)/83.8(0.14)} & \cellcolor{gray!34}{\textbf{80.2}} \\
\ding{51} & \ding{51} & \ding{55} & \ding{51} & \ding{55} & \ding{51} & \cellcolor{gray!45}{80.7(0.14)} & \cellcolor{gray!44}{86.3(0.15)} & \cellcolor{gray!43}{82.7(0.13)/80.8(0.12)} & \cellcolor{gray!44}{74.4(0.12)} & \cellcolor{gray!44}{89.4(0.17)} & \cellcolor{gray!44}{84.2(0.15)/88.5(0.17)} & \cellcolor{gray!44}{60.2(0.05)} & \cellcolor{gray!43}{85.3(0.14)/85.4(0.14)} & \cellcolor{gray!44}{\textbf{81.6}} \\
%\ding{51} & \ding{51} & \ding{55} & \ding{51} & \ding{51} & \ding{55} & \cellcolor{gray!38}{79.6(0.12)} & \cellcolor{gray!41}{85.8(0.13)} & \cellcolor{gray!35}{81.4(0.14)/79.7(0.13)} & \cellcolor{gray!29}{72.3(0.10)} & \cellcolor{gray!39}{88.6(0.18)} & \cellcolor{gray!39}{83.4(0.17)/87.2(0.17)} & \cellcolor{gray!43}{60.1(0.05)} & \cellcolor{gray!39}{84.6(0.12)/84.7(0.17)} & \cellcolor{gray!37}{\textbf{80.7}} \\
%\ding{51} & \ding{51} & \ding{55} & \ding{51} & \ding{51} & \ding{51} & \cellcolor{gray!39}{79.8(0.15)} & \cellcolor{gray!37}{85.2(0.16)} & \cellcolor{gray!36}{81.6(0.14)/80.0(0.13)} & \cellcolor{gray!31}{72.6(0.08)} & \cellcolor{gray!40}{88.7(0.13)} & \cellcolor{gray!41}{83.7(0.18)/88.3(0.21)} & \cellcolor{gray!41}{59.9(0.05)} & \cellcolor{gray!39}{84.6(0.16)/84.8(0.14)} & \cellcolor{gray!39}{\textbf{80.8}} \\
\ding{51} & \ding{51} & \ding{51} & \ding{55} & \ding{55} & \ding{55} & \cellcolor{gray!38}{79.7(0.11)} & \cellcolor{gray!38}{85.3(0.18)} & \cellcolor{gray!36}{81.6(0.14)/80.0(0.13)} & \cellcolor{gray!27}{72.1(0.10)} & \cellcolor{gray!35}{88.0(0.17)} & \cellcolor{gray!40}{83.6(0.14)/88.2(0.17)} & \cellcolor{gray!36}{59.4(0.06)} & \cellcolor{gray!42}{85.2(0.13)/84.4(0.13)} & \cellcolor{gray!37}{\textbf{80.7}} \\
% \ding{51} & \ding{51} & \ding{51} & \ding{55} & \ding{55} & \ding{51} & \cellcolor{gray!35}{79.3(0.12)} & \cellcolor{gray!36}{85.1(0.17)} & \cellcolor{gray!34}{81.3(0.12)/80.0(0.13)} & \cellcolor{gray!27}{72.2(0.11)} & \cellcolor{gray!36}{88.1(0.16)} & \cellcolor{gray!42}{83.9(0.14)/87.2(0.16)} & \cellcolor{gray!41}{59.9(0.06)} & \cellcolor{gray!39}{84.6(0.15)/84.4(0.13)} & \cellcolor{gray!36}{\textbf{80.5}} \\
\ding{51} & \ding{51} & \ding{51} & \ding{55} & \ding{51} & \ding{55} & \cellcolor{gray!45}{80.7(0.13)} & \cellcolor{gray!44}{86.3(0.14)} & \cellcolor{gray!45}{82.9(0.13)/80.9(0.13)} & \cellcolor{gray!38}{73.6(0.10)} & \cellcolor{gray!48}{90.2(0.18)} & \cellcolor{gray!45}{84.3(0.15)/88.6(0.18)} & \cellcolor{gray!44}{60.3(0.06)} & \cellcolor{gray!45}{85.5(0.13)/85.3(0.15)} & \cellcolor{gray!44}{\textbf{81.7}} \\
% \ding{51} & \ding{51} & \ding{51} & \ding{55} & \ding{51} & \ding{51} & \cellcolor{gray!45}{80.8(0.11)} & \cellcolor{gray!46}{86.6(0.18)} & \cellcolor{gray!46}{83.1(0.13)/81.2(0.13)} & \cellcolor{gray!38}{73.6(0.11)} & \cellcolor{gray!48}{90.2(0.15)} & \cellcolor{gray!47}{84.7(0.16)/88.6(0.18)} & \cellcolor{gray!46}{60.5(0.05)} & \cellcolor{gray!45}{85.6(0.17)/85.4(0.14)} & \cellcolor{gray!45}{\textbf{81.8}} \\
\ding{51} & \ding{51} & \ding{51} & \ding{51} & \ding{55} & \ding{55} & \cellcolor{gray!34}{79.1(0.13)} & \cellcolor{gray!40}{85.7(0.14)} & \cellcolor{gray!39}{81.9(0.13)/80.1(0.15)} & \cellcolor{gray!25}{71.9(0.10)} & \cellcolor{gray!37}{88.3(0.15)} & \cellcolor{gray!37}{83.1(0.16)/88.0(0.14)} & \cellcolor{gray!37}{59.4(0.05)} & \cellcolor{gray!37}{84.3(0.17)/84.3(0.12)} & \cellcolor{gray!37}{\textbf{80.6}} \\
\ding{51} & \ding{51} & \ding{51} & \ding{51} & \ding{55} & \ding{51} & \cellcolor{gray!50}{\textbf{\underline{81.4(0.14)}}} & \cellcolor{gray!48}{86.9(0.15)} & \cellcolor{gray!50}{83.6(0.16)/81.5(0.16)} & \cellcolor{gray!40}{73.9(0.12)} & \cellcolor{gray!49}{90.2(0.18)} & \cellcolor{gray!50}{\textbf{\underline{85.1(0.12)}}/89.5(0.15)} & \cellcolor{gray!50}{60.8(0.05)} & \cellcolor{gray!48}{86.0(0.15)/85.8(0.14)} & \cellcolor{gray!48}{\textbf{82.3}} \\
\ding{51} & \ding{51} & \ding{51} & \ding{51} & \ding{51} & \ding{55} & \cellcolor{gray!47}{80.9(0.12)} & \cellcolor{gray!47}{86.8(0.16)} & \cellcolor{gray!48}{83.3(0.15)/81.2(0.14)} & \cellcolor{gray!39}{73.8(0.11)} & \cellcolor{gray!48}{90.1(0.19)} & \cellcolor{gray!47}{84.7(0.14)/89.3(0.18)} & \cellcolor{gray!48}{\textbf{\underline{60.9(0.06)}}} & \cellcolor{gray!47}{85.9(0.14)/85.9(0.18)} & \cellcolor{gray!47}{\textbf{82.1}} \\
\ding{51} & \ding{51} & \ding{51} & \ding{51} & \ding{51} & \ding{51} & \cellcolor{gray!50}{\textbf{\underline{81.4(0.13)}}} & \cellcolor{gray!50}{\textbf{\underline{87.3(0.17)}}} & \cellcolor{gray!50}{\textbf{\underline{83.7(0.13)/81.8(0.13)}}} & \cellcolor{gray!50}{\textbf{\underline{75.3(0.12)}}} & \cellcolor{gray!50}{\textbf{\underline{90.4(0.19)}}} & \cellcolor{gray!50}{\textbf{\underline{85.1(0.16)/89.6(0.20)}}} & \cellcolor{gray!50}{60.7(0.05)} & \cellcolor{gray!50}{\textbf{\underline{86.4(0.13)/86.3(0.14)}}} & \cellcolor{gray!50}{\textbf{\textbf{\underline{82.6}}}} \\
\bottomrule \bottomrule
\end{tabular}
\begin{tablenotes}
\item[1] Determined \ac{PT} using Alg.~\ref{alg:methods}.
\item[2] \ac{HWA} finetuned weights evaluated without noise in FP.
\end{tablenotes}
\end{threeparttable}
\vspace{-1em}
\end{table*}

\subsubsection{Channel-wise Conductance Range Learning}
can be performed by introducing a learnable parameter, $\eta \in (0, 1]$, which scales the computed output current of each \ac{BL}. By representing inputs and weights as normalized values, i.e., $\in [-1, 1]$, after training is completed, the conductance range for each channel (\ac{BL}) can be computed, $\eta G_{max}$, for which weights can be scaled to.

\subsection{Post-Training Optimization Methods}
\ac{PT} optimization methods for input and conductance ranges are described in Alg.~\ref{alg:methods}. These can be performed in addition to, or without, input range and/or channel-wise conductance range learning. For conductance range optimization, $N=2$ was considered. This was determined empirically.

\section{Results}~\label{sec:results}
First, to establish a baseline, after initial training, RoBERTA was retrained, as described in Section~\ref{sec:hwa_retraining}, for each task without any non-idealities (see Table~\ref{table:architecture}).
To evaluate each configuration after \ac{HWA} finetuning, programming noise was applied, and for multiple (10) instances, the corresponding task score was determined at different points in time. It is noted there is a significant drop in accuracy between the baseline and most accurate configuration for each task, as the RoBERTA model was not retrained using the original corpora.
%. We attribute this to the lack of retraining the RoBERTA model using the original corpora. We refer the reader to~\cite{raschHardwareawareTrainingLargescale2023a} for a more comprehensive \ac{HWA} training study.

The temporal dependence of the score for one of the tasks (RTE) is evaluated in Fig.~\ref{fig:time_scale} for four different configurations.  
To investigate the effectiveness of both \ac{PT} optimization methods, and to compare them to \ac{HWA} training configurations where these parameters are learned, we report the scores for a number of combinations of conductance and input range configurations at $t=$1h in Table~\ref{table:main_results}. For tasks other than RTE, the temporal dependence did not differ substantially.

\section{Discussion and Outlook}
\textbf{Post-training optimization methods result in significant accuracy recovery:} When \ac{PT} optimization methods were employed, a significant recovery in accuracy was observed when only weight clipping and I/O quantization was performed during training (on average, 5.3\%). Additionally, compared to when both input and conductance ranges were learned, the average score decreased by only 0.1\% when they were optimized \ac{PT}, compared to 5.3\% when they were not. 

\textbf{Further optimizing learned parameters can improve accuracy:} When both input and conductance ranges were learned and optimized \ac{PT}, the average score increased by 0.9\%. When optimizing these parameters \ac{PT}, more detailed and hardware-specific aspects can be captured without compromising training scalability.

\textbf{Hyper-parameter tuning:}
By decreasing the number of hyper-parameters which affect \ac{HWA} training performance, in addition to reducing the computational complexity, especially when hyper-parameters are not tuned, training stability and generalizability is increased.  

\textbf{Regular calibration:} Due to temporal dynamics, as observed in Fig.~\ref{fig:time_scale}, for current \ac{AIMC} accelerators, some degree of regular calibration is required to retain accuracy over sustained periods of operation. \ac{PT} optimization methods, such as input range optimization, can be performed, which do not require the costly reprogramming of devices.

\section{Conclusion}~\label{sec:conclusion}
In this paper, we improved the accuracy of \ac{AIMC} accelerators \ac{PT} using two novel optimization methods. The effectiveness of these methods were demonstrated empirically using fine-tuned RoBERTA models, and eight GLUE tasks with realistic hardware simulations. Using both methods, the average score across all tasks was increased from 76.3 to 81.6\%. Regular calibration experiments, investigation of more types of sizes of models, and hardware experiments form the basis of future work.

\bibliographystyle{IEEEtran}
\bibliography{References}

\end{document}